\documentclass[12pt]{iopart}

\usepackage{graphicx}
\usepackage{dcolumn}
\usepackage{bm}
\usepackage[colorlinks=true]{hyperref}
\usepackage{url}
\usepackage{color}
\usepackage{braket}
\usepackage{textcomp}
\usepackage{listings}
\usepackage{bbold}
\usepackage{float}

\expandafter\let\csname equation*\endcsname\relax
\expandafter\let\csname endequation*\endcsname\relax
\usepackage{amsmath}
\usepackage[capitalise]{cleveref}

\begin{document}

\title{A quantum-classical cloud platform optimized for variational hybrid algorithms}
\author{Peter J.\ Karalekas, Nikolas A.\ Tezak\footnote{Current address: OpenAI, 3180 18th St, San Francisco, CA 94110 USA}, Eric C.\ Peterson,\\Colm A.\ Ryan, Marcus P.\ da Silva\footnote{Current address: Microsoft Quantum, One Microsoft Way, Redmond, WA 98052 USA}, and Robert S.\ Smith}
\address{Rigetti Computing, 2919 Seventh Street, Berkeley, CA 94710 USA}

\date{\today}

\begin{abstract}
In order to support near-term applications of quantum computing, a new compute paradigm has emerged---the quantum-classical cloud---in which quantum computers (QPUs) work in tandem with classical computers (CPUs) via a shared cloud infrastructure. In this work, we enumerate the architectural requirements of a quantum-classical cloud platform,
and present a framework for benchmarking its runtime performance.
In addition, we walk through two platform-level enhancements, parametric compilation and active qubit reset, that specifically optimize a quantum-classical architecture to support variational hybrid algorithms (VHAs), the most promising applications of near-term quantum hardware. Finally, we show that integrating these two features into the Rigetti Quantum Cloud Services (QCS) platform results in considerable improvements to the latencies that govern algorithm runtime.
\end{abstract}

\maketitle

\section{Introduction}

The first experimental realizations of quantum algorithms date back to over a decade ago \cite{IkeNMRProcessor, KwiatOpticsProcessor, GuideIonProcessor, DiCarloSCProcessor}, but in the last three years quantum computing has rapidly transitioned from a field of scientific research to a full-fledged technology industry. The recent demonstration of \textit{quantum supremacy} over classical computing \cite{GoogleSupremacy} is a considerable milestone, but there is still much progress to be made on the road to solving real-world problems with quantum computers and achieving \textit{quantum advantage}. Improving the error rates of quantum devices \cite{Ballance2016, HongEntanglingGate} and ultimately reaching the regime of \textit{fault tolerance} \cite{PreskillSupremacy} is necessary for unlocking the most powerful known applications of quantum computers. At the same time, the industry has increased its focus on finding ways to solve valuable problems using the noisy intermediate-scale quantum (NISQ) processors that are currently available \cite{PreskillNISQ}. 

The desire to provide the research community with access to scarce quantum hardware in order to shorten the path to quantum advantage resulted in the development of a new compute architecture---the \textit{quantum cloud}. As part of this architecture, the concept of an Internet-accessible data center has been extended to include quantum devices. Infrastructure for the quantum cloud requires a slew of new specialized hardware, for example, dilution refrigerators to house superconducting qubits and racks of microwave instruments to control them. To build the quantum cloud, some developers of quantum computers have pivoted to being \textit{full-stack}, using in-house infrastructure to offer cloud-based access to their quantum devices (e.g.\ IBM Quantum Experience). In addition, some traditional cloud providers have begun to add quantum backends through strategic hardware-software partnerships.

The first iteration of quantum cloud offerings employed a hybrid cloud model \cite{NISTCloudComputing}, in which users of the service submitted quantum programs using a web API to a queue hosted by the public cloud (e.g.\ Amazon Web Services). Then, a server colocated with a quantum processor would periodically pull jobs off of the queue, execute them, and return results back to the user. This approach was effective in offering worldwide, public access to quantum resources, but suffers in terms of runtime efficiency due to the overhead of network connections and the use of a shared queue. In addition, the traditional web API model fails to capitalize on or adapt to any properties specific to using a quantum device for computation.

In particular, the most promising approach to effectively using near-term quantum devices is through \textit{variational hybrid algorithms} (VHAs) \cite{McCleanTheoryVQE} which employ a quantum-classical architecture, essentially leveraging the quantum computer as a co-processor alongside a powerful classical computer. These algorithms have been applied to areas such as combinatorial optimization \cite{FarhiQAOA, RigettiClustering}, quantum chemistry \cite{PeruzzoPhotonicVQE, GoogleXmonVQE}, and machine learning \cite{RigettiQKS}, and numerous proposals for applications of the variational method continue to arise with increasing frequency \cite{Bravo-PietroVQLS, GuillaumeVQT}. However, VHAs require a tight coupling between quantum and classical resources, and using a cloud-hosted queue is slow with respect to the scale of quantum operations, especially on a superconducting device \cite{OliverSCStateOfPlay}. In addition, a quantum cloud architecture must be specifically optimized in order to efficiently support the variational model of execution. In this work, we investigate architectural bottlenecks of this new \textit{quantum-classical cloud}, and provide a benchmarking framework to analyze its runtime performance. We then use the benchmark to quantify the dramatic reduction in latency achieved by the Rigetti Quantum Cloud Services (QCS) platform via the implementation of specialized techniques for quantum program compilation and qubit register reset.

\section{Runtime bottlenecks in the quantum-classical cloud}

The job of a quantum cloud platform is to ingest programs written in a backend-independent high-level quantum programming language \cite{SmithQuil, CrossQASM}, compile them into a platform-specific representation, run them on an available quantum device, and return the results to the user. Specifically, a quantum cloud platform has four essential components:

\begin{enumerate}
    \item An apparatus that houses the physical objects that act as qubits (e.g.\ an optical table and trapping system for ions or neutral atoms).
    \item A \textit{control system} containing instruments for manipulating that apparatus in order to drive the desired evolution and read out qubit measurement results.
    \item An \textit{executor} that orchestrates the control system to run quantum programs and return measurement results to the user.
    \item A \textit{compiler} that takes in quantum programs and produces instrument binaries for the executor.
\end{enumerate}

\begin{figure}
    \centering
    \includegraphics[width=\textwidth]{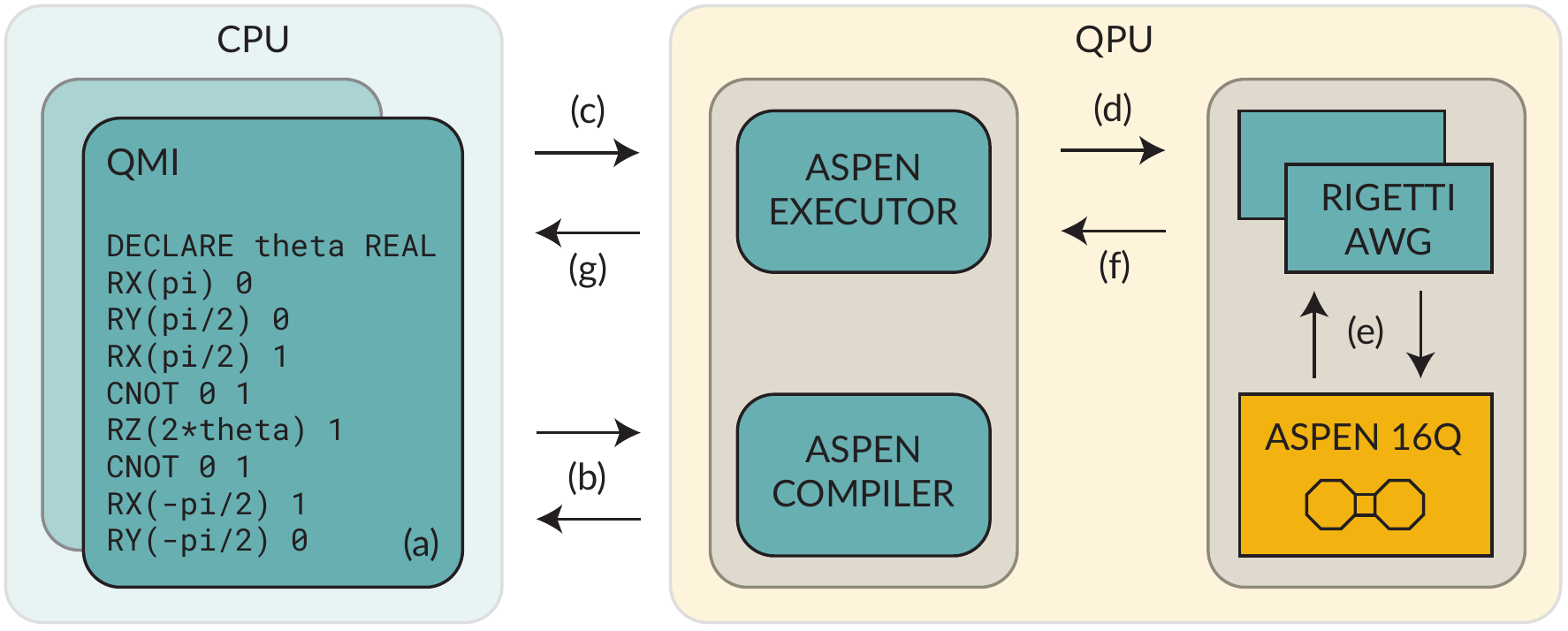}
    \caption{
    Schematic diagram of the Quantum Cloud Services (QCS) data center. The classical compute (CPU) is connected over the network to a collection of quantum processors (QPUs), which are composed of a classical host computer, control system rack, dilution refrigerator, and quantum integrated circuit (QuIC).
    (a) Users write quantum programs using Quil on the QMI, which is located inside the data center.
    (b) The quantum program is sent to the compiler. 
    (c) The binaries produced in (b) are sent to the executor.
    (d) The executor loads the binaries onto a collection of Rigetti AWGs.
    (e) Rigetti AWGs send microwave pulses into the dilution refrigerator containing the Aspen-4 16Q device. 
    (f--g) Bitstring results are returned to the QMI.
    }
    \label{fig:QCS-Architecture}
\end{figure}

To be categorized as quantum-classical cloud, a platform must also include access to classical compute resources. Depending on the particular qubit implementation used by the platform, the CPU-QPU interaction could become the largest bottleneck in the variational model of execution. For example, when using superconducting qubits (with gate times in the tens of nanoseconds) the CPU and QPU should be physically colocated in order to enable a low-latency link between the user and the quantum device. Although colocating user and compute is not a new concept in cloud computing in general, it has yet to take hold broadly in quantum computing, and drastically reduces overhead in VHAs. For Rigetti QCS, which uses superconducting qubits as the backend, users interact with the QPU via a preconfigured development environment called the quantum machine image (QMI) (\cref{fig:QCS-Architecture}a). The QMI is a virtual machine running on a classical compute cluster located inside the Rigetti \textit{quantum data center} in Berkeley, CA, and contains the Forest SDK for building applications using the quantum instruction language Quil \cite{SmithQuil}. Once written, quantum programs are sent for compilation into pulse-level instructions (\cref{fig:QCS-Architecture}b). The information that encodes this gate-to-pulse mapping is contained within a calibration database, and is updated whenever the system drifts out of specification. The binaries that are returned by the compiler are then sent to the executor (\cref{fig:QCS-Architecture}c), which loads them onto a collection of Rigetti-custom microwave arbitrary waveform generators (AWGs) and receivers, and triggers the instruments to begin execution (\cref{fig:QCS-Architecture}d). The Rigetti AWGs then send microwave pulses into a dilution refrigerator to manipulate and read out the state of the Aspen-4 16-qubit quantum processor (\cref{fig:QCS-Architecture}e). Finally, the bitstrings resulting from readout are returned to the user's compute environment (\cref{fig:QCS-Architecture}f--g) for processing and analysis.

In the variational model of execution, one seeks to minimize an objective function that is expensive to compute on classical hardware, by embedding it as a subroutine on a quantum computer. To begin, a classical computer takes an initial guess and instructs a quantum computer to perform a predetermined computation. Then, from the output statistics of sampling the quantum program many times, a classical optimizer running on the CPU updates the QPU instructions for the next round of iteration. Depending on the difficulty of the problem, and the quality of the QPU, this will repeat many times before finding a potential solution \cite{McCleanBarrenPlateaus}. Thus, the structure of a variational hybrid algorithm is broken into two nested loops: an outer \textit{variational iteration loop}, and an inner \textit{quantum execution loop}. We denote each roundtrip completion of the variational iteration loop as an \textit{iteration step}, and each roundtrip completion of the quantum execution loop as a \textit{shot}. Each iteration step includes communicating with the QPU, waiting for it to return results, and using an optimizer to decide what to run on the QPU next. Each shot is defined by running a single program of quantum instructions and getting back a single bitstring of measurement outcomes. This structure gives us a way to think about the different components that contribute to hybrid algorithm runtime---they either occur once per step in the variational iteration loop, or once per shot in the quantum execution loop. In order to improve the performance of QCS and optimize it for the variational model of execution, we began by constructing a latency budget for each of the nested loops (\cref{table:Latency-Budget}). 

The components of the per-step latency budget are instrument initialization, network communication, and compilation, which is the largest contributor by at least one order of magnitude. The contribution from instrument initialization has already been substantially reduced by using custom control hardware. In addition, the runtime of a single shot is split between gates (for manipulating and entangling qubit states), readout (for measuring and extracting bitstring outcomes), and qubit reset. Qubit reset can be performed passively, by waiting for all qubits to relax to their ground states. However, this relaxation process is the same decoherence process that determines $T_1$, one of the metrics for qubit lifetime. Therefore, this \textit{passive qubit reset} time will only increase as qubit performance improves \cite{RigettiFabT1, GyensisZeroPi}, and it already dominates the per-shot latency budget. Thus, from examining the two latency budgets, compilation and qubit reset are the areas to which improvements would have the largest impact on algorithm runtime.

\begin{table}[H]
\centering
\begin{tabular}{|c|c|}
\multicolumn{2}{c}{Per-step latency budget} \\ [0.5ex] 
\hline
Task & Time \\ 
\hline
Compilation & $200\,\mathrm{ms}$ \\
AWG load \& arm & $8\,\mathrm{ms}$ \\
AWG trigger & $10\,\mathrm{ms}$ \\
Network comms & $5\,\mathrm{ms}$  \\
\hline
\end{tabular}
\quad
\quad
\begin{tabular}{|c|c|}
\multicolumn{2}{c}{Per-shot latency budget} \\ [0.5ex] 
\hline
Task & Time \\ 
\hline
Single-qubit gates & $60\,\mathrm{ns}$ \\
Two-qubit gates & $300\,\mathrm{ns}$ \\
Readout \& capture & $2\,\mu\mathrm{s}$ \\
Passive reset & $100\,\mu\mathrm{s}$  \\
\hline
\end{tabular}
\caption{Per-step and per-shot latency budgets for the Aspen-4 QPU via QCS, rounded to one significant figure. The per-step budget was collected by running Max-Cut QAOA \cite{FarhiQAOA} on two qubits, and measuring the completion time for each of the components. The per-shot budget contains the average gate and readout times for Aspen-4 from the calibration database. The reset time is five times the longest $T_1$ time measured on Aspen-4 (see \cref{fig:Active-Reset} for details). The compilation task is further broken down into two components: the first which converts arbitrary quantum programs into ones that use the gateset and topology of the target quantum device (about $150\,\mathrm{ms}$) \cite{PetersonQuil}; and the second which converts this \textit{nativized} program into pulse-level instrument binaries (about $50\,\mathrm{ms}$).}
\label{table:Latency-Budget}
\end{table}

\section{Optimizing for the variational execution model}

Having identified compilation and qubit reset as potential bottlenecks for a quantum-classical cloud architecture, we walk through specific enhancements made to mitigate their contributions to the latency budgets of Rigetti's quantum cloud platform.

\subsection{Parametric compilation}

The underlying quantum program for a variational algorithm is parametric, meaning that from iteration step to iteration step, the sequence of instructions is static and only the instruction arguments change. Thus, a specialized compiler that preserves this \textit{parametric ansatz} structure can leverage it to improve runtime. To build such a compiler, we need to understand the physical implementation of a high-level quantum program in order to determine which instructions are easy to modify parametrically at runtime. For superconducting quantum processors (and many other QPU implementations), qubits are controlled by shaped radio frequency (RF) or microwave pulses typically generated by an AWG. The pulse amplitude, duration, and phase control the rotation angle and axis effected on the qubits. Typically the phase of the pulse is determined by an abstract rotating \textit{reference frame} set to a frequency determined by the physics of the qubits. $Z$-rotations of a qubit then correspond to instantaneous reference frame update events that change the phase of subsequent microwave pulses. Without loss of generality, we can directly relate each parameter of a variational algorithm to $Z$-rotations or phase updates for one or more qubit reference frames. Previously, the arguments to these \texttt{SHIFT-PHASE} \cite{QuilRepo} operations were provided at compile-time, and hardcoded in the instrument binary. Thus, for every step in a VHA, new instrument binaries would need to be compiled, such that these arguments could be updated.

\begin{figure}
    \centering
    \includegraphics[width=\textwidth]{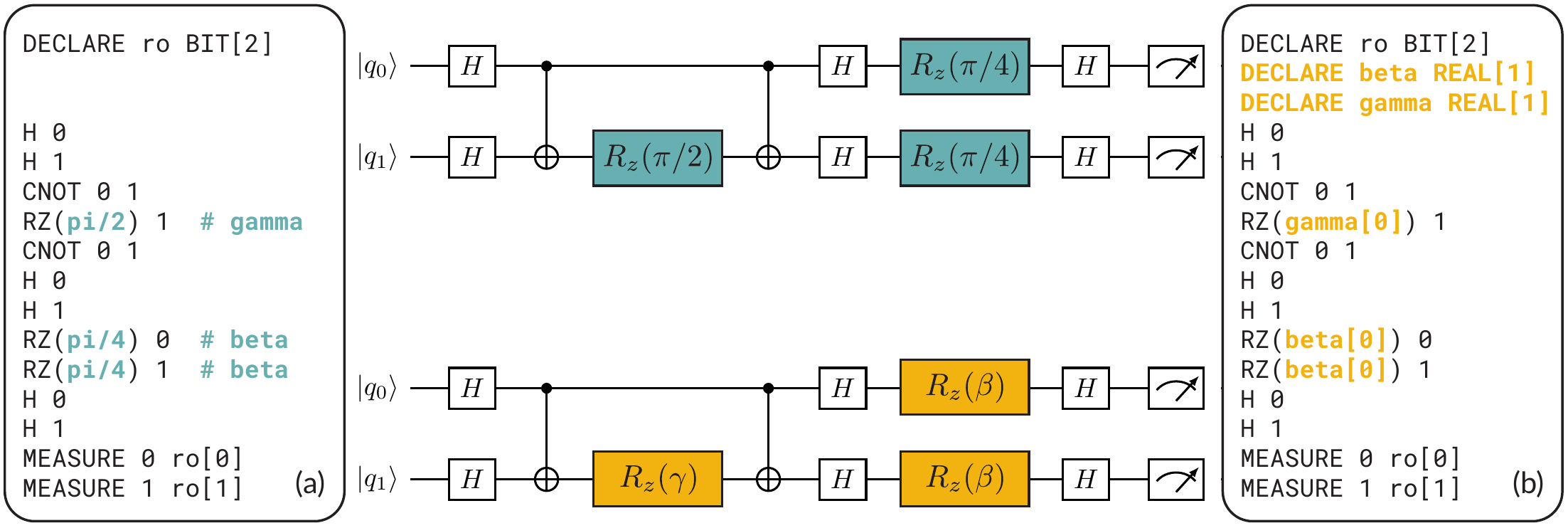}
    \caption{
    Quantum programs, written in Quil, for Max-Cut QAOA on two qubits, and their corresponding circuit diagrams. (a, top) The program, using a specific set of values for variational parameters ($\beta$, $\gamma$). These show up as arguments to \texttt{RZ} gates, with $\beta = \pi/4$ and $\gamma = \pi/2$. Thus, every iteration step, the program itself has to be updated with new variational parameters and re-compiled before execution. (b, bottom) The program, taking advantage of parametric compilation. Now, rather than providing values for $\beta$ and $\gamma$ when writing the program, we instead initialize real-valued classical memory registers \texttt{beta} and \texttt{gamma} using Quil's \texttt{DECLARE} syntax, and use them as arguments for the \texttt{RZ} gates. This allows for their assignment to be deferred from compile-time to run-time, and eliminates the need for re-compilation between variational iteration steps.
    }
    \label{fig:Parametric-Compilation}
\end{figure}

To circumvent this need for re-compiling, we implemented a feature called \textit{parametric compilation}. Like a standard executable file, each instrument binary includes a header and a collection of sections \cite{ELFStandard}. The \textit{instruction memory} section contains executable code, and the \textit{waveform memory} section contains pulse shapes that are referenced by the instruction memory. Rather than having \texttt{SHIFT-PHASE} instructions use a static data field as an argument, each instrument binary now additionally has a \textit{data memory} section which instructions can access by reference. Like classical shared memory, the data memory section can also be updated at runtime by an external process in order to change the effect of the binary \cite{POSIXStandard}. The  data memory section layout of a particular binary is prescribed by Quil's \texttt{DECLARE} syntax (\cref{fig:Parametric-Compilation}), which can be used to initialize named memory registers of various data types (\texttt{BIT}, \texttt{OCTET}, \texttt{INTEGER}, and \texttt{REAL}). Once initialized, the memory registers can be provided as arguments to parameterized gates such as \texttt{RZ}. When these gates are compiled (\cref{fig:QCS-Architecture}b) they become \texttt{SHIFT-PHASE} operations that reference entries in an empty data memory section. Upon execution, the user provides these  \textit{parametric binaries} along with a map of memory register assignments, for example, variational parameters for the current iteration step of a VHA (\cref{fig:QCS-Architecture}c). Then, the executor fills in the data memory section using the memory map to produce \textit{patched binaries} (\cref{fig:QCS-Architecture}d). For each step in a VHA, these binaries can be re-patched with a new set of variational parameters, and thus the variational iteration loop can repeat until termination without the need for compilation. For details on how to implement a variational algorithm using parametric compilation, see \labelcref{appA}.

\subsection{Active qubit reset}

Rather than waiting for qubits to passively reset, we can implement a protocol called \textit{active qubit reset} that sets all qubits to their ground states at the beginning of a computation. This has been previously demonstrated by controllably transferring qubit excited state population to a system with a much faster decay rate (such as the readout resonators) \cite{EggerActiveReset, MagnardActiveReset, GoogleSupremacy}, or by performing a measurement and then quickly feeding back to conditionally apply an $X$ gate \cite{RisteActiveReset, RyanBBNHardware}. But, for this protocol to be useful, the conditional control flow needs to happen on a timescale comparable to the quantum gates, and it cannot introduce significant error into proceeding computations.

We chose to implement this classical feedback loop on our platform because it additionally unlocks more complex feedback/feedforward circuits that take advantage of mid-circuit measurements. To support this, the compilation toolchain can propagate control flow structures down to the hardware pulse sequencers. Coupled with the ability to rapidly broadcast qubit measurement results across the control system, active reset is not a special-case operation but instead a simple \texttt{if}-\texttt{then}-\texttt{else} control flow branching off of qubit measurement results. By providing the \texttt{RESET} instruction at the top of a Quil program, users can signal to the compilation toolchain that they would like to enable active qubit reset. When ingested by the compiler, the \texttt{RESET} directive is translated to control-flow graphs (CFGs) \cite{DragonBook} for each qubit that encapsulate the branching and looping structure of the active reset protocol. 

\begin{figure}
    \centering
    \includegraphics[width=\textwidth]{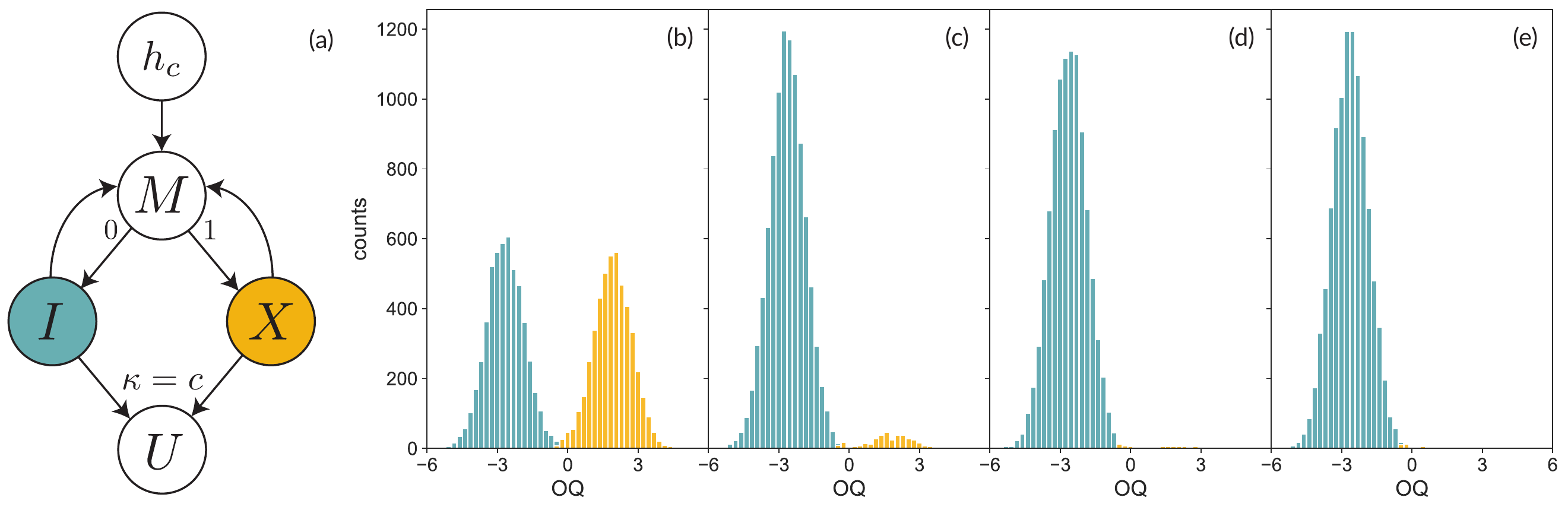}
    \caption{
    Active qubit reset on a single qubit. (a) The \texttt{RESET} control-flow graph, containing loop and branch constructs. The program starts at header block $h_c$, which is initialized with $c$, the number of active reset rounds to perform. It then jumps to measurement block $M$ and increments the feedback counter $\kappa$. If the measurement result is a 0, the program then jumps to idle block $I$; otherwise, it jumps to feedback block $X$. After $I$ or $X$ the program jumps back to $M$ unless $\kappa = c$. In that case it jumps to $U$, the main body block. (b--e) Optimal quadrature histograms of IQ signal data from three successive rounds of active reset (c--e) on a single Aspen-4 qubit starting from an equal superposition input state (b). On Aspen-4, $T_1$ times range from 10 to $20\,\mu\mathrm{s}$. Thus, in order to achieve a reset fidelity $\mathcal{F} > 99\,\%$ in the passive reset case, we must wait for $5 T_1 = 100\,\mu\mathrm{s}$ between shots (as $1 - e^{-5} > 0.99$). If we instead use active qubit reset, we can perform three rounds of measurement and feedback on the order of $10\,\mu \mathrm{s}$. By the third round (e), we achieve a reset fidelity of $\mathcal{F} = 99.6\,\%$.
    }
    \label{fig:Active-Reset}
\end{figure}

Control-flow graphs are composed of nodes called \textit{basic blocks}, each of which includes a sequence of instructions without branching and up to two directed edges, or \textit{jump targets}, to follow once the instructions are complete. If a block has two jump targets, it also contains a conditional expression for choosing between them. Each single-qubit \texttt{RESET} CFG contains four basic blocks---a header block, a measurement block, an idle block, and a feedback block (\cref{fig:Active-Reset}a). The program starts at the header block, which initializes a counter to track the number of active reset rounds performed, and then jumps to the measurement block and increments the counter. The measurement block contains readout and capture instructions for the qubit, and conditionally jumps to either the idle or feedback blocks dependent on the bit result produced by the measurement. If the result is 1, the qubit is in the excited state, and therefore the feedback block containing the $X$ gate program is executed. Otherwise, if the result is 0, the program jumps to the idle block. After either the idle or feedback block is completed, the program then jumps to the measurement block again if the counter is less than the number of reset rounds requested. After each single-qubit program traverses its active reset CFG, the program jumps to the first basic block of the main body quantum circuit and proceeds until completion.

The time required to perform the active reset protocol is dependent on the underlying architecture of a QPU. For Aspen-4, we require three rounds of feedback to achieve a reset performance comparable to that of passive qubit reset (\cref{fig:Active-Reset}b--e). All active reset sequences must be completed before the main body program begins, and the readout operations on Aspen-4 are in the $2$ to $3\,\mu\mathrm{s}$ range. Combined with a broadcast and feedback latency of around $1\,\mu\mathrm{s}$, this results in an active reset time of $9$ to $12\,\mu\mathrm{s}$, approximately a tenfold improvement over passive reset times.

\section{A volumetric framework for benchmarking runtime}

A useful metric for determining the runtime performance of a quantum-classical cloud platform is QPU latency, which tells us how long the CPU has to wait between requesting execution on a quantum device and receiving back the results. Generically, this latency depends on the number of qubits involved in the computation, the number of shots requested, and the program being run. We can define a function $\mathcal{T}(m,n,\mathcal{P})$ which returns QPU latency and takes a number of qubits $m$, a number of shots $n$, and a function $\mathcal{P}$ that produces a family of quantum programs for a given number of qubits. Some examples of $\mathcal{P}$ are \verb|GHZ_LINE|, which produces the straight-line program to create a GHZ state on $m$ qubits \cite{GHZState}, and \verb|MAXCUTQAOA_COMPLETE|, which produces a QAOA program to solve Max-Cut on a fully-connected graph with $m$ nodes \cite{FarhiQAOA}. In addition, by sweeping $n$ and measuring $\mathcal{T}$ with fixed $m$ and $\mathcal{P}$, we can determine two asymptotic latencies of a platform for a particular family of quantum programs and number of qubits: the single-shot limit ($n=1$) tells us the \textit{variational step latency} ($\mathcal{T}_V$) and the scaling of the latency in the limit of large $n$ tells us the \textit{quantum shot latency} ($\mathcal{T}_Q$). More formally, we assume that the total latency $\mathcal{T}$ is straightforwardly related to the two components in the following manner
\begin{equation}
    \mathcal{T}(m, n, \mathcal{P}) = \mathcal{T}_{V}(m, \mathcal{P}) + n \mathcal{T}_{Q}(m, \mathcal{P}).
\end{equation}

However, $\mathcal{T}_V$ and $\mathcal{T}_Q$ are still functions that depend on the number of qubits and the requested quantum program, and therefore need further specification to be an effective cross-platform benchmark. Naively choosing $m = 1$ and $\mathcal{P}$ to be circuits containing only \texttt{MEASURE} instructions makes it trivial to experimentally benchmark the system. By sweeping the number of shots, measuring latency, and fitting the data to a linear model, the resulting latency-axis intercept gives $\mathcal{T}_V(1, \texttt{MEASURE})$ and the resulting slope gives $\mathcal{T}_Q(1,\texttt{MEASURE})$. These numbers well-encapsulate some parts of the performance of a particular architecture, but fail to capture how the architecture fares as the number of qubits increases or as the program complexity changes. For example, potential pitfalls such as long gate times or a poorly scaling control system initialization routine are entirely omitted by the benchmark. But, simply setting $m$ to be the maximum device size available on a particular platform is equally misleading, as current NISQ devices often have error rates such that they cannot produce useful entanglement across the entire QPU.

\begin{figure}
    \centering
    \includegraphics[width=\textwidth]{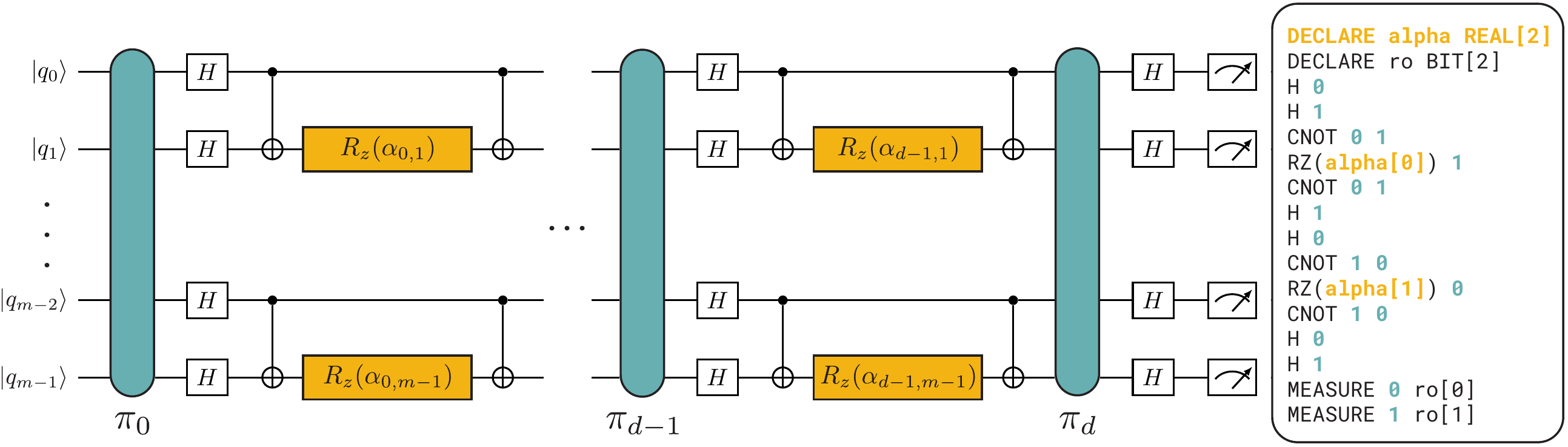}
    \caption{
    Random phase gadget (RPG) family of volumetric circuits, used for benchmarking runtime on a quantum-classical cloud platform. (left) The qubits and layers are indexed starting at zero, and the angle values $\alpha_{i,j}$ for qubit $j$ and layer $i$ are chosen at random. Although the circuit family could be defined for an arbitrary number of layers $d$ and qubits $m$, we choose $m = d = \log_{2}{V_Q}$ in order to determine computationally relevant latencies. It is important to note, however, that this choice is arbitrary and only meant to simplify the benchmark. As in quantum volume, if the number of qubits is odd, the bottom qubit line has no gates. To benchmark a number of qubits $m$, we choose a set of permutations $\{\pi_i\}$, run the resulting circuit $r$ times, and compute the average runtime. For each run, we randomize all the $\alpha$ values and collect $n$ shots. This effectively emulates a VHA \cite{CrooksQAOA}, as the permutations are fixed ahead of time, and only the phase gadgets themselves change. Then, repeating this entire process for multiple permutation sets ensures that we get a good estimate of the average runtime for a particular number of qubits.  (right) Example Quil circuit from $\texttt{RPG}(2)$, meaning $m = d = 2$, and using parametric compilation to defer the assignment of $\alpha$.}
    \label{fig:Phase-Gadget}
\end{figure}

Although there is no unanimously supported benchmark for QPU performance within the community (and there may never be), $\log$ quantum volume \cite{CrossQuantumVolume, MollVariational} ($\log_{2}{V_Q}$) has been proposed as a reasonable near-term metric for the number of qubits that can be meaningfully used in a computation. We are interested in something that is similar to quantum volume (so that the choice of number of qubits remains relevant), but more appropriate for the near-term VHAs. Variational algorithms contain structures known as \textit{phase gadgets} \cite{CowtanPhaseGadgets}, which are \texttt{RZ} gates sandwiched between \texttt{CNOT}s. These structures are often the cornerstone of parametric-ansatz-style programs, and therefore we propose using a volumetric family of circuits~\cite{RobinVolumetrics} that we call \textit{random phase gadgets} (RPG) to benchmark algorithm runtime. The RPG circuit family incorporates the permutation aspect of quantum volume for exercising connectivity, parallelism, and gateset~\cite{PetersonPolytopes}, but replaces the random 2Q unitaries with phase gadgets that have \texttt{RZ} gates with randomly chosen arguments (\cref{fig:Phase-Gadget}). In addition, each permutation is followed by a layer of Hadamard gates on all qubits, to make it more difficult to compile away the phase gadgets. Setting $m = \log_{2}{V_Q}$ and $\mathcal{P} = \texttt{RPG}$ gives us the \textit{computationally relevant} step ($T_V$) and shot ($T_Q$) latency of a QPU
\begin{align}
T_{V} &= \mathcal{T}_{V}(\log_{2}{V_Q}, \texttt{RPG}), & T_{Q} &= \mathcal{T}_{Q}(\log_{2}{V_Q}, \texttt{RPG}).
\end{align}
Fitting the resulting runtime data to the linear model $T(n) = T_{V} + n T_{Q}$ then allows us to easily estimate variational algorithm runtimes using the computationally relevant QPU latency for a particular device available on a quantum cloud platform.

\section{QPU latency results on Quantum Cloud Services}

\begin{figure}
    \centering
    \includegraphics[width=\textwidth]{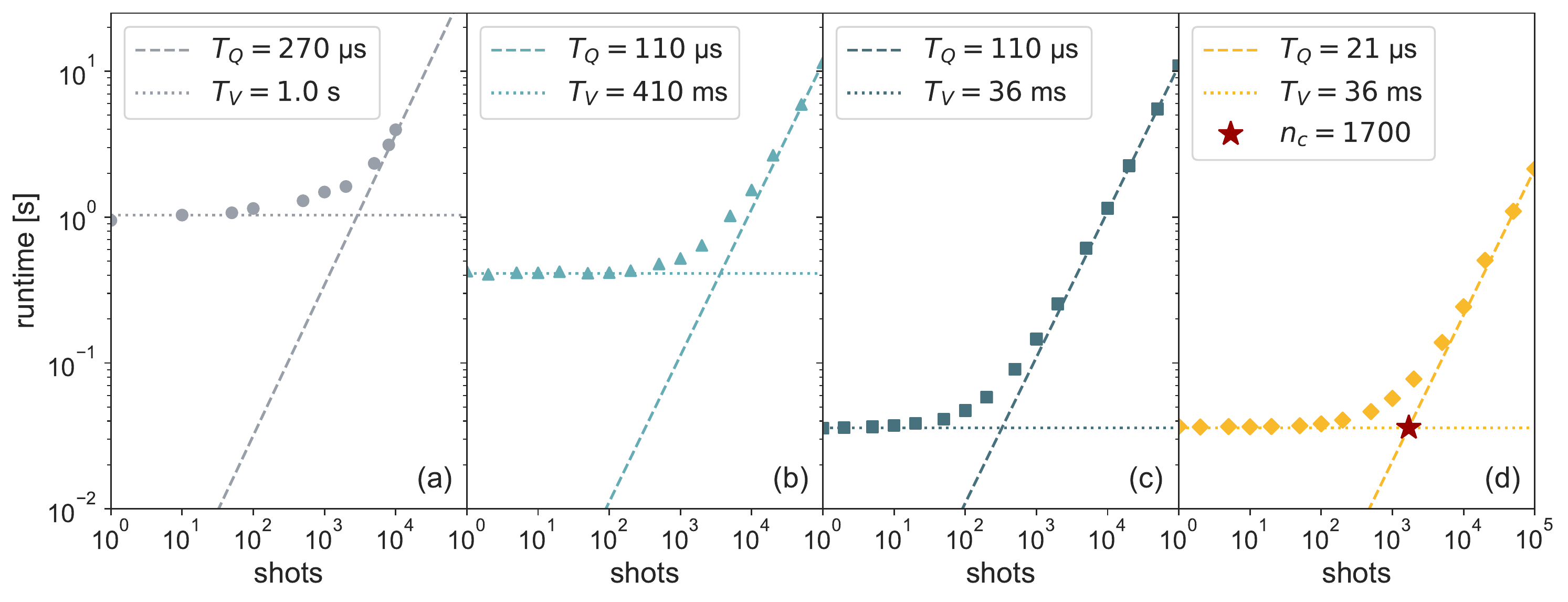}
    \caption{
    Benchmarking QPU latency for Rigetti's quantum cloud, plotted on a log-log scale to aid in visualization of the asymptotic behavior.
    (a) Median runtime data from various quantum programs run by external users via the Forest Web API, across the top ten numbers of shots (1, 10, 50, 100, 500, 1000, 2000, 5000, 8000, 10,000) used on that version of the platform.
    (b) Median runtime data collected according to the framework in Section 4 via the Rigetti Quantum Cloud Services (QCS) platform, using the Aspen-4 QPU which has $\log_{2}{V_Q} = 3$. The QCS data is taken for 1, 2, 5, 10, 20, 50, 100, 200, 500, 1000, 2000, 5000, 10,000, 20,000, 50,000, and 100,000 shots, and each median is extracted from 100 runs of the benchmark with a fixed permutation set. (c) Median runtime data collected as in (b), but with parametric compilation enabled.
    (d) Median runtime data collected as in (c), but with active qubit reset enabled. For this optimal platform configuration, we additionally note that the \textit{critical shot number} ($n_c$), which is the
    turning point between $T_V$-dominated QPU latency and $T_Q$-dominated QPU latency, occurs at the 1700-shot mark.}
    \label{fig:QPU-Latency}
\end{figure}

With a volumetric framework for benchmarking runtime, we calculate $T_V$ and $T_Q$ for four different versions of the Rigetti quantum cloud platform. To demonstrate the initial performance improvements resulting from simply colocating CPU and QPU, we first analyzed runtime data from over 851,000 quantum programs run by external users on the Acorn and Agave QPUs via the Forest Web API, the initial version of Rigetti's platform (\cref{fig:QPU-Latency}a). The Forest Web API used the first-generation model of quantum cloud access, routing each job through a queue on Amazon Web Services (AWS), which resulted in considerable latencies. Because the Forest Web API programs were run by external users, there is no expectation that the average composition of these programs would match the composition of the benchmark programs described in Section 4. However, on average the Forest Web API programs used 3 qubits, 2 \texttt{CZ} gates, and 14 \texttt{RX} gates. A program from $\texttt{RPG}(3)$, upon compilation to the native gateset available on Acorn and Agave, would have more of both gates, and therefore would take longer to execute. This, combined with the fact that the log quantum volume for Aspen-4 is $\log_{2}{V_Q} = 3$, makes it reasonable to compare the data from the Forest Web API and QCS, and in fact skews the comparison in favor of the former. Thus, using median runtime data from Forest Web API, we calculate $T_Q = 270\,\mu\mathrm{s}$ and $T_V = 1.0\,\mathrm{s}$. If we instead use the benchmarking framework from Section 4 and collect data via QCS's colocated architecture, $T_V$ drops to $410\,\mathrm{ms}$ and $T_Q$ to $110\,\mu\mathrm{s}$ for the Aspen-4 QPU (\cref{fig:QPU-Latency}b). Using parametric compilation (\cref{fig:QPU-Latency}c), we can remove the compile step from our runtime calculations, resulting in an improvement for small numbers of shots ($T_V$ drops to $36\,\mathrm{ms}$). For higher numbers of shots, passive reset times still dwarf the constant improvement from parametric compilation. Finally, by enabling active qubit reset (\cref{fig:QPU-Latency}d), we get an additional reduction in latency within the quantum execution loop. Thus, in this optimal configuration of the platform, $T_V = 36\,\mathrm{ms}$ and $T_Q = 21\,\mathrm{\mu}\mathrm{s}$, resulting in greater than $27\times$ and $12\times$ improvements, respectively, over the latencies of the first-generation access model.

\section{Conclusions}

Quantum Cloud Services may be the first instance of a quantum-classical cloud platform, but this architectural paradigm will become increasingly common as the industry continues to progress toward useful applications of quantum computers. Error rates and qubit count are well-known to be important system benchmarks within the field, but as more and more hardware providers begin to offer access to quantum resources over the cloud, the latencies that govern this access and the resulting application runtimes will also be critical considerations for platform performance. Addressing these latencies requires approaching system bottlenecks with an interdisciplinary \textit{quantum software engineering} mindset, bridging the knowledge bases of classical and quantum computing. We have shown that colocation, parametric compilation, and active qubit reset provide considerable improvements over the first-generation of quantum cloud offerings, but they are just a few of the many potential platform optimizations for accelerating industry progress and enabling the achievement of quantum advantage.

\ack

RSS designed the language constructs that support parametric compilation. PJK, ECP, NAT, and RSS built the parametric compilation toolchain. NAT implemented the software for expressing active qubit reset as a control-flow graph. CAR coordinated the integration of active reset into the QCS platform. PJK, ECP, and CAR formulated the benchmark, and PJK collected and analyzed all runtime data. PJK and MPS architected the framework for running VHAs using parametric compilation. RSS supervised the QCS effort. PJK, CAR, and MPS wrote the manuscript and prepared the figures\footnote{All of the plotting and analysis from the paper can be recreated using the supplementary Jupyter notebooks and datasets \cite{QCSPaperSupplement}, which are also available on the notebook hosting service \href{https://mybinder.org/v2/gh/rigetti/qcs-paper/master?urlpath=lab/tree/Welcome.ipynb}{Binder}.}.

This work was funded by Rigetti \& Co Inc., dba Rigetti Computing. We thank the Rigetti quantum software team for providing tooling support, the Rigetti fabrication team for manufacturing the device, the Rigetti technical operations team for fridge maintenance, the Rigetti cryogenic hardware team for providing the chip packaging, the Rigetti control systems and embedded software teams for creating the Rigetti AWG control system, and the Rigetti quantum engineering team for building the infrastructure for automated QPU bringup and recalibration.

In addition, the authors would like to specifically thank Lauren E.\ Capelluto, Steven Heidel, and Anthony M.\ Polloreno for their critical contributions to the QPU compiler toolchain, Glenn E.\ Jones, Rodney F.\ Sinclair, and Blake R.\ Johnson for their work on designing a control system capable of supporting active reset, Alexander D. Hill, Joseph A.\ Valery, and Alexa N. Staley for their management of QPU performance and deployment, Zachary P.\ Beane, Adam D.\ Lynch, Nicolas J.\ Ochem, and Christopher B.\ Osborn for their orchestration of the QMI compute infrastructure, E.\ Schuyler Fried, Diego Scarabelli, and Prasahnt Sivarajah for their crucial work on experimental bringup of active qubit reset, Joshua Combes, Kyle V.\ Gulshen, and Nicholas C.\ Rubin for building the readout error mitigation tools that enable the implementation of VHAs, Matthew P.\ Harrigan and William J.\ Zeng for their leadership in designing a quantum programming interface to QCS, and Matthew J.\ Reagor for insightful discussions on modeling and benchmarking QPU latency.

\appendix

\section{Implementing hybrid algorithms with parametric compilation}
\label{appA}

The structure of quantum programs for variational hybrid algorithms can be segmented into two parts \cite{McCaskeyBenchmarkVQE}: parameterized ansatz preparation, and measurement in a variety of multi-qubit Pauli bases. Parametric compilation handles not only the parameterized ansatz component of a VHA, but can also be used to encapsulate measurement in arbitrary bases and symmetrization of readout error---all in a single quantum program. This allows for any VHA to be expressed via a single parametric binary with \texttt{RZ} arguments that are provided and patched at runtime. To show how this is possible, we walk through three examples using pyQuil \cite{PyQuilZenodo}, the Python library for writing and executing Quil programs.

\subsection{Readout error symmetrization and mitigation}

\begin{figure}
    \centering
    \includegraphics[width=\textwidth]{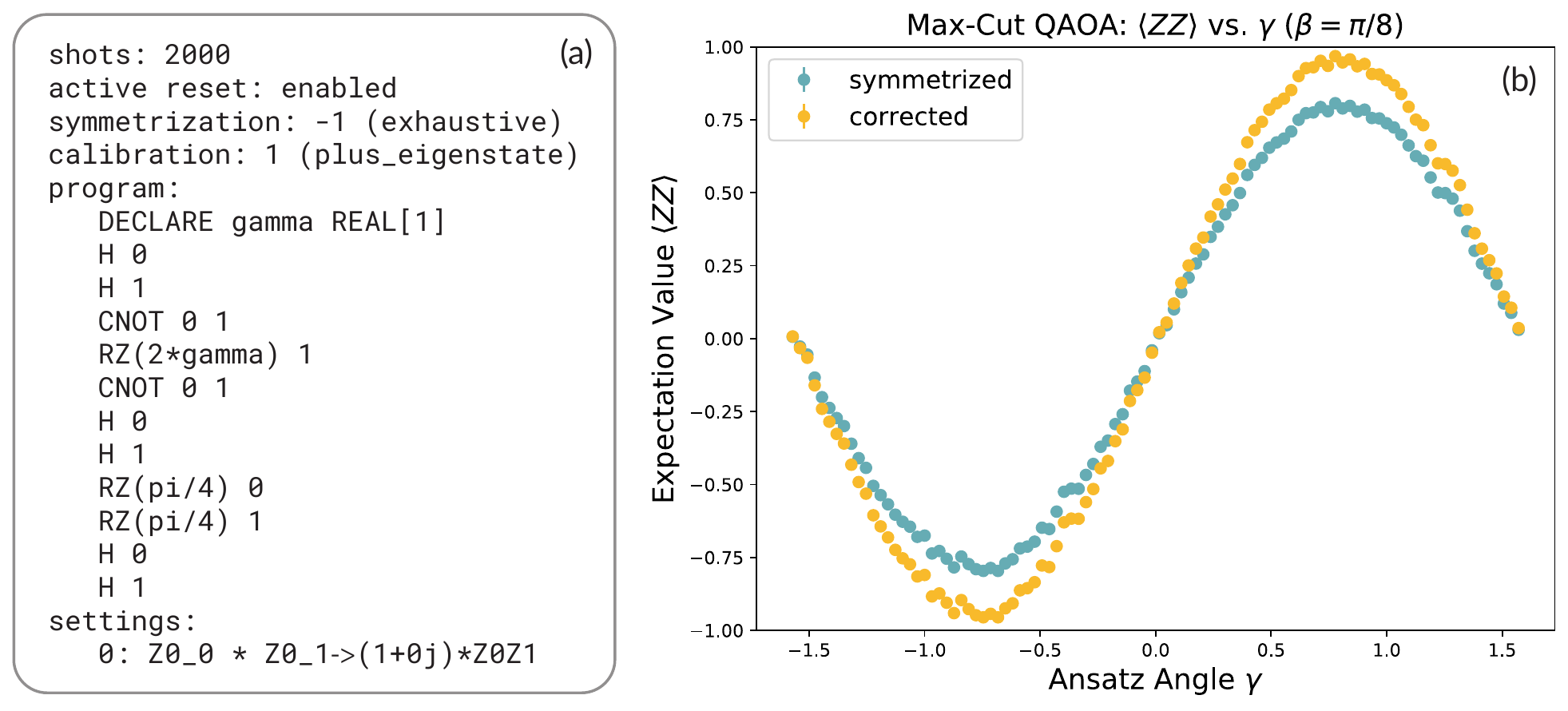}
    \caption{
    Running 2Q Max-Cut QAOA for $p=1$ and fixed $\beta$ on qubits 1 and 2 of the Aspen-4 QPU using parametric compilation and pyQuil's \texttt{Experiment} framework. (a) Max-Cut QAOA expressed as an \texttt{Experiment} object. The \texttt{program} section contains the Quil program equivalent to the Max-Cut QAOA ansatz for $p = 1$ and $\beta = \pi/8$. The ansatz is composed of three parts: an initial $\ket{++}$ state preparation, the mixer unitary $U_m(\beta) = e^{-i\beta (X_0 + X_1)}$, and the cost unitary $U_c(\gamma) = e^{-i\gamma Z_0 Z_1}$. As the Max-Cut QAOA cost function uses only $ZZ$ expectation values, the \texttt{settings} section contains only one entry. (b) Results from running Max-Cut QAOA for fixed $\beta$ using qubits 1 and 2 on the Aspen-4 QPU, sweeping $\gamma$ for 100 values in $[-\pi/2, \pi/2]$. The \textit{symmetrized} data points are collected using exhaustive readout symmetrization, but without correcting for imperfect readout. The \textit{corrected} data points are the symmetrized data points, but rescaled using the results from pyQuil's \texttt{calibration} method on the $ZZ$ expectation value for the qubits used in the \texttt{Experiment}.}
    \label{fig:Symmetrization}
\end{figure}

We can describe measurement imperfections in terms of a \textit{confusion matrix} 
\textbf{M}, which describes probabilities of reported measurement outcomes 
conditioned on the values of the true state of the qubits. For the case of a single 
qubit, the confusion matrix can be written as
\begin{equation}
\textbf{M} = \begin{pmatrix}
           \text{Pr}(0|0) & \text{Pr}(1|0) \\
            \text{Pr}(0|1) & \text{Pr}(1|1) \\
            \end{pmatrix}
         = \begin{pmatrix}
            1 - \epsilon_0 & \epsilon_1\\
            \epsilon_0 & 1 - \epsilon_1 \\
            \end{pmatrix}.
\end{equation}
While generally $\textbf{M}$ will not be symmetric with respect to the exchange of 0s 
and 1s, one can enforce such symmetry by flipping bits before measurement and 
subsequently flipping the measurement outcomes. This {\em symmetrization} procedure 
corresponds to a simple form of {\em twirling}~\cite{Bennet1996,Bartlett2007}, and in principle can be performed by randomly choosing which bits to 
flip independently each time a measurement occurs\footnote{More sophisticated 
approaches to symmetrization can be taken by making use of orthogonal arrays, as 
implemented in {\tt forest-benchmarking}~\cite{ForestBenchmarking}, but a detailed 
description of those techniques is beyond the scope of this article.}.
When averaging  over all the shots, this results in a new \textit{effective} confusion matrix $\textbf{M}'$, with $\epsilon_0$ and $\epsilon_1$ replaced by $\epsilon'$, 
which is the arithmetic mean of the two. This can be implemented in Quil (for a single qubit with index 0) by replacing the measurement section of a program with:
\begin{center}
\begin{tabular}{l}
\verb|DECLARE symmetrization REAL[1]|\\
\verb|DECLARE ro BIT[1]|\\
\verb|RX(symmetrization[0]) 0|\\
\verb|MEASURE 0 ro[0]|\\
\end{tabular}
\end{center}
During the execution of the program, the \texttt{symmetrization} memory region assignments $0$ and $\pi$ must be provided, and the results from the $\pi$ assignment must be flipped (i.e. \texttt{XOR}ing the results with the bit value of 1).

It is straightforward to check that the effect of symmetric measurement errors is to scale the expectation value of a Pauli observable by factor that depends on the error rates. This immediately suggests how to mitigate the effect of these errors: first characterize the error rates of the symmetrized readout, then rescale the observed expectation values accordingly. The first step is what we call {\em readout calibration}, while the second step is what we call {\em readout error mitigation}. Since errors due to the single-qubit rotations associated with mapping one Pauli observable to another are orders of magnitude smaller than measurement errors, readout calibration can focus on the expectation of tensor products of $Z$ observables only. Ideally the expectation of any of these tensor products would be $+1$ for the ground state (the $\ket{00\cdots0}$ state), but for symmetrized readout error the expectation value will be $\lambda$. Simply dividing the expectation value of this same observable for other states by $\lambda$ corrects for the bias introduced by the measurement errors. Note that, in general, even with symmetrization, a different $\lambda$ for each different tensor product will be necessary, due to qubit-dependent readout signal-to-noise ratio (SNR) and potential correlations in the readout errors. Moreover, while the bias in the estimation of the expectation value is removed, the uncertainty is increased (as $|\lambda|<1$), so this procedure is not scalable~\cite{Ryan2015}. That being said, it is remarkably effective for small numbers of qubits.

Rather than handling symmetrization manually, we can use pyQuil's \texttt{Experiment} framework. When defining an \texttt{Experiment} in pyQuil, two arguments are required---a \texttt{program} which defines the main body quantum circuit, and a list of \texttt{settings} which specify the different state preparations and measurements that we want to wrap around the main program. In addition, the \texttt{Experiment} object also specifies the number of shots to take, whether or not to perform active qubit reset, and how to symmetrize and calibrate readout. To show how these features are used, we implement readout symmetrization and correction for a 2Q Max-Cut QAOA program (\cref{fig:Symmetrization}).

\subsection{Bell state tomography}

\begin{figure}
    \centering
    \includegraphics[width=0.7\textwidth]{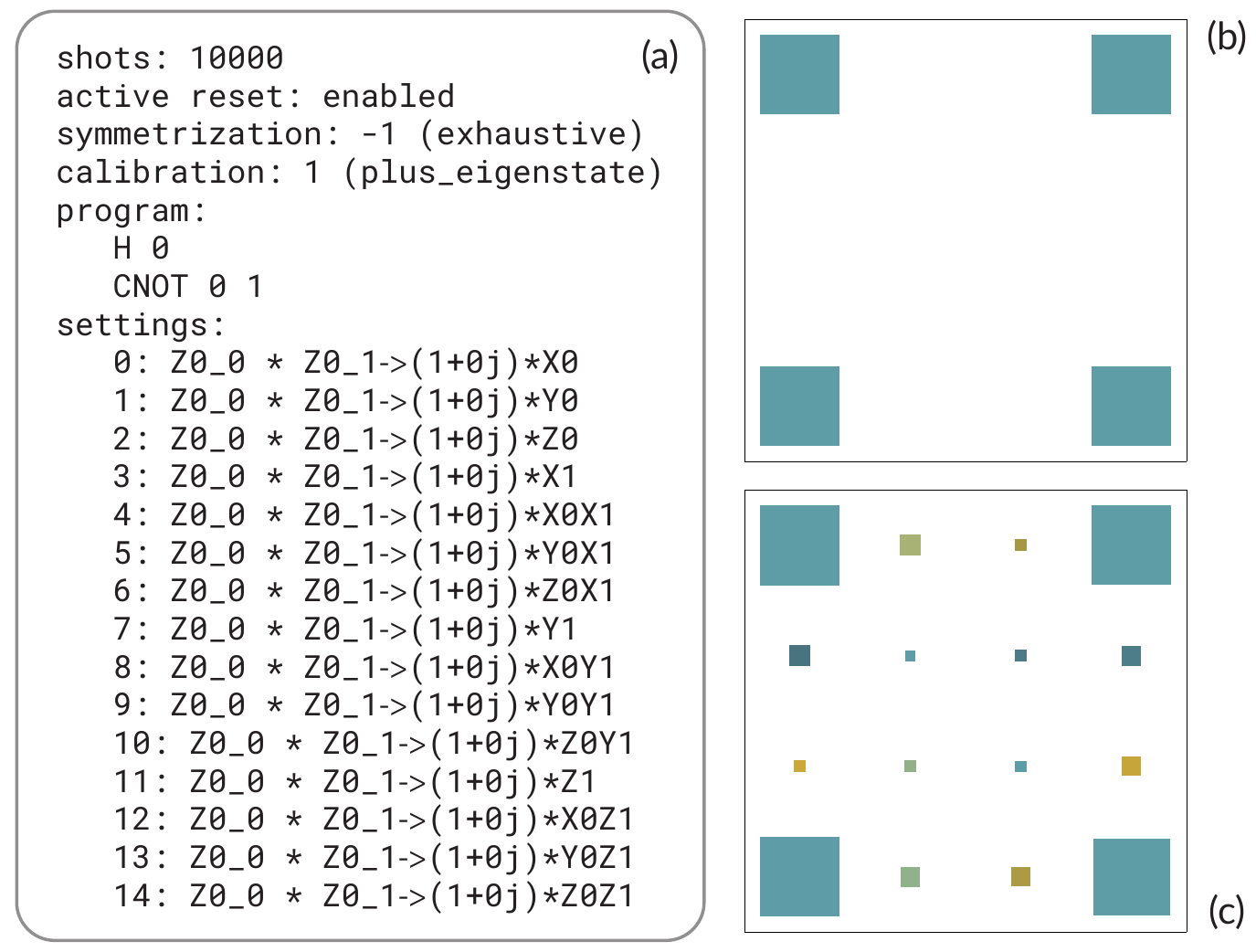}
    \caption{
    Running Bell state tomography on qubits 1 and 2 of the Aspen-4 QPU using parametric compilation and pyQuil's \texttt{Experiment} framework. (a) Bell state tomography, expressed as an \texttt{Experiment} object. The \texttt{program} section contains the gates required to generate the Bell state $\ket{\Phi^+} = \frac{1}{\sqrt{2}}(\ket{00} + \ket{11})$. The \texttt{settings} section contains the 15 different non-identity Pauli measurements required to tomograph a 2Q state, generated by the software library \texttt{forest-benchmarking} \cite{ForestBenchmarking}. (b) Hinton plot \cite{Matplotlib, SciPy} of the ideal density matrix $\rho$ as defined by the state $\ket{\Phi^+}$. (c) Hinton plot of the estimated density matrix $\rho_{\textrm{est}}$, extracted from readout-corrected experimental data using the linear inversion method \cite{WoodThesis}. We calculate a Bell state fidelity of $\mathcal{F}_{\Phi^+} = 99.35\,\%$ by comparing $\rho$ and $\rho_{\textrm{est}}$ using the \texttt{fidelity} function from \texttt{forest-benchmarking}.}
    \label{fig:Tomography}
\end{figure}

An arbitrary single-qubit gate $U$ can always be decomposed into what is in essence an Euler angle decomposition~\cite{MikeAndIke}
\begin{align}
    U(\alpha, \beta, \gamma) &= R_z(\gamma) R_x(-\pi/2) R_z(\beta) R_x(\pi/2) R_z(\alpha).
\end{align}
Although our QPUs can only perform measurements in the $z$ basis ($M_z$), performing single-qubit rotations prior to measurement allows us to effectively measure in a different basis
\begin{align}
    M_x &= M_z R_y(-\pi/2) = M_z U(0, -\pi/2, 0),\\
    M_y &= M_z R_x(\pi/2) = M_z U(\pi/2, \pi/2, -\pi/2).
\end{align}
Thus, by pre-pending a qubit measurement with an Euler-decomposed single-qubit gate containing $R_z$ arguments that can be changed at runtime, we can perform a collection of different measurements with a single parametric binary. In Quil, this is accomplished for a single qubit by adding the following snippet before the measurement block:
\begin{center}
\begin{tabular}{l}
\verb|DECLARE measurement_alpha REAL[1]|\\
\verb|DECLARE measurement_beta REAL[1]|\\
\verb|DECLARE measurement_gamma REAL[1]|\\
\verb|RZ(measurement_alpha[0]) 0|\\
\verb|RX(pi/2) 0|\\
\verb|RZ(measurement_beta[0]) 0|\\
\verb|RX(-pi/2) 0|\\
\verb|RZ(measurement_gamma[0]) 0|\\
\end{tabular}
\end{center}
Additionally, this can be combined with readout symmetrization, allowing for any desired observable to be symmetrized, calibrated, and corrected. To show how measurement bases can be changed parametrically, we run Bell state tomography using a single binary (\cref{fig:Tomography}).

\subsection{The variational quantum eigensolver}

\begin{figure}
    \centering
    \includegraphics[width=\textwidth]{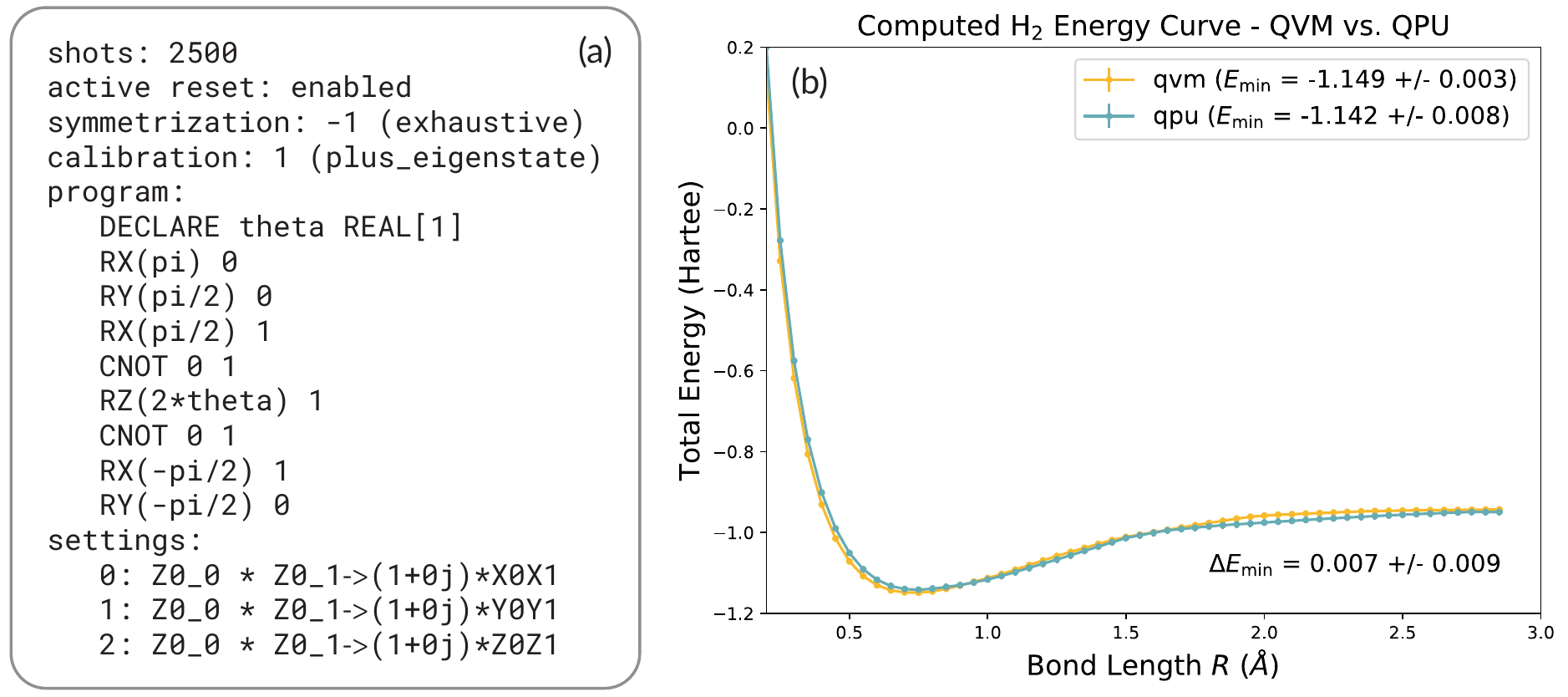}
    \caption{
    Running the variational quantum eigensolver to compute the ground state energy of the hydrogen molecule using parametric compilation and pyQuil's \texttt{Experiment} framework. (a) The VQE H\textsubscript{2} algorithm, expressed as an \texttt{Experiment} object. The \texttt{program} section contains the parameterized UCC ansatz $U(\theta)$ (\cref{eq:Ansatz}), and the \texttt{settings} section contains the three measurement bases (expressed as Pauli strings) required to estimate the expectation value of the Hamiltonian. The Hamiltonian (\cref{eq:Hamiltonian}) also contains single-qubit $Z$ terms, but they can be determined from measurement outcomes of the $ZZ$ setting. (b) Results from running the VQE H\textsubscript{2} \texttt{Experiment} on the Quantum Virtual Machine (QVM) \cite{RigettiQVM} as well as qubits 1 and 2 on the Aspen-4 QPU. We use the table of Hamiltonian coefficients from the appendix of O'Malley et al. to convert our readout-corrected Pauli expectations into expectation values for the H\textsubscript{2} Hamiltonian at different bond lengths. At the minimum-energy bond length $R_{\textrm{min}} = 0.750\,\textup{\AA}$ we measure an energy difference $\Delta E_{\textrm{min}} = 8 \pm 9\,\mathrm{mHa}$. To demonstrate the framework, we simply scanned 250 values of $\theta$ in the range $[-\pi/2, \pi/2]$, but the latency numbers would not differ if we instead used an optimizer to update $\theta$ each step. Collecting the data took $248\,\textrm{s}$, compared to $266\,\textrm{s}$ as predicted by $T(2500) = 88.5\,\textrm{ms}$ repeated for 4 symmetrization settings, 3 measurement bases, and 250 values of $\theta$ (for a total of 3000 executions on the QPU). This slight improvement over the prediction is consistent, as $T(n)$ from Section 5 is calculated for a higher number of qubits (3) and number of two-qubit gates than are used in this VHA.}
    \label{fig:VQE}
\end{figure}

Finally, we combine the techniques from the previous two examples in order to run a full variational algorithm using a single parametric binary. The \textit{variational quantum eigensolver} (VQE) \cite{PeruzzoPhotonicVQE}, which is one of the leading VHAs for applications in quantum chemistry, can be used to compute the ground state energy of the hydrogen molecule (H\textsubscript{2}). To do so, VQE employs a classical minimization routine in which the objective function is evaluated on the QPU. The procedure begins by preparing a parameterized ansatz wavefunction $\ket{\Psi(\theta)}$ using an initial guess for the variational parameter $\theta$. The parameterized ansatz wavefunction can be chosen to be composed of the unitary coupled cluster (UCC) ansatz $U(\theta)$ and the Hartree-Fock (HF) reference state $\ket{\Phi}$ \cite{GoogleXmonVQE}, giving
\begin{align}
    \ket{\Psi(\theta)} = U(\theta)\ket{\Phi} = e^{-i \theta X_0 Y_1}\ket{01}.
\label{eq:Ansatz}
\end{align}
Then, the procedure continues by computing the expectation value of the H\textsubscript{2} Hamiltonian, which can be formulated as a sum of multi-qubit Pauli-basis expectation values with coefficients
\begin{align}
    H = g_0 \mathbb{1} + g_1 Z_0 + g_2 Z_1 + g_3 Z_0 Z_1 + g_4 Y_0 Y_1 + g_5 X_0 X_1.
\label{eq:Hamiltonian}
\end{align}
Finally, the expectation value $\braket{H}$ is fed to the minimizer to choose the variational parameter $\theta$ for the next round of iteration, and this repeats until convergence. Following the protocol in O'Malley et al., we use parametric compilation and pyQuil's \texttt{Experiment} framework to simulate the ground state energy of the H\textsubscript{2} molecule with VQE (\cref{fig:VQE}).

\section*{References}
\bibliographystyle{phunsrt}
\bibliography{qcs}

\end{document}